# A Roadmap to Address Burnout in the Cybersecurity Profession: Outcomes from a Multifaceted Workshop

Ann Rangarajan[1], Calvin Nobles[2], Josiah Dykstra[3], Margaret Cunningham[4], Nikki Robinson[5], Tammie Hollis[6], Celeste Lyn Paul[7], and Charles Gulotta[8]

[1] Department of Information Technology and Management, Illinois Institute of Technology,
arangarajan@iit.edu
[2] School of Cybersecurity and Information Technology, University of Maryland Global Campus,
calvin.nobles@umgc.edu
[3] Trail of Bits, Work done while at NSA,
josiah.dykstra@trailofbits.com
[4] Wethos AI,
margaret@thehumanconsultant.com
[5] Capitol Technology University,
dr.nikki.robinson@gmail.com
[6] Muma College of Business, University of South Florida,
nrt@usf.edu
[7] National Security Agency, Work done while at Veros Technologies,
clpaul@nsa.gov
[8] National Security Agency,
csgulot@nsa.gov

**Abstract.** This paper addresses the critical issue of burnout among cybersecurity professionals, a growing concern that threatens the effectiveness of digital defense systems. As the industry faces a significant attrition crisis, with nearly 46% of cybersecurity leaders contemplating departure from their roles, it is imperative to explore the causes and consequences of burnout through a socio-technical lens. These challenges were discussed by experts from academia and industry in a multi-disciplinary workshop at the 26th International Conference on Human-Computer Interaction to address broad antecedents of burnout, manifestation and its consequences among cybersecurity professionals, as well as programs to mitigate impacts from burnout. Central to the analysis is an empirical study of former National Security Agency (NSA) tactical cyber operators. This paper presents key insights in the following areas based on discussions in the workshop: lessons for public and private sectors from the NSA study, a comparative review of addressing burnout in the healthcare profession. It also outlines a roadmap for future collaborative research, thereby informing interdisciplinary studies.

# 1      Introduction

Cybersecurity is a rapidly evolving socio-technical problem, confronted by increasingly sophisticated and pervasive threats. As organizations and nations confront these challenges, an emerging critical issue threatens the very foundation of digital defense: burnout among cybersecurity professionals [1]. The cybersecurity industry is facing an attrition crisis, with studies indicating that up to 46% of cybersecurity leaders are considering leaving their roles [2]. Not only does this alarming trend impact individual organizations, but it also poses a significant risk to national and global security.

Burnout amongst cybersecurity professionals is increasingly acknowledged in both industry settings and academic literature [1, 3, 4, 5]. The World Health Organization has classified burnout as an occupation-related syndrome [6] that can result in an increased error rate, leading to cybersecurity risk with significant associated and financial or reputational impact on the organization. Notwithstanding this risk, it increases employee turnover, decreases productivity, causes excessive stress, decreases motivation, and increases fatigue [4, 7]

Experts across academia and industry discussed these challenges in a multidisciplinary workshop at the 26th International Conference on Human-Computer Interaction. As a collaborative outcome of the workshop, this paper presents the *What, Why,* and *How* of the burnout phenomenon in the cybersecurity profession. By drawing insights from comparative analyses, empirical studies, and interdisciplinary perspectives, we propose a roadmap for addressing this time-sensitive concern.

This article benefits both industry practitioners and cybersecurity scholars alike. By examining this nascent topic with a socio-technical lens we provide a deeper understanding of the problem to develop early-stage interventions at individual, organizational, and societal levels. By synthesizing insights from diverse sources and proposing concrete strategies for mitigation, we aim to catalyze a shift in how the industry approaches workforce sustainability. As cyber threats continue to evolve, ensuring the long-term resilience of our human cybersecurity resources is paramount. Our findings provide a foundation towards this endeavor, calling for collective action from researchers, organizations, and policymakers to safeguard our digital systems and the individuals who defend them

## 1.1      Evolution of Burnout in the Cybersecurity Profession

Early contextual studies on burnout focused on the Information Technology (IT) professional [8, 9, 10, 11, 12]. Table 1 summarizes demographics of IT job roles referenced in such empirical studies. It is evident that cybersecurity related roles were not explicitly identified.

**Table 1.** IT professional roles referenced in burnout-related empirical studies.

| Reference to Burnout Study | IT Roles |
|---|---|
| Pawlowski et al. [10] | System administrator, network planning engineer, IT project manager, IT management consultant, Vice president of IT services |
| Kim [16] | IT management, networks, PC end-user support, system development & integration, technical service & operations |
| Rutner et al. [15] | Database administrator, manager, programmer/analyst, support |
| Shih et al. [13] | Programmer, system analyst, project manager, R&D, engineer, data processing, network administration, line manager, CIO |
| Brooks et al. [14] | Business or systems analyst, database administrator, Software developer, project manager, team lead, IS/IT director or manager, technical support staff (hardware and software) |
| | Run-in Heading in Bold. Text follows |

## 1.2 Bridging Gaps in Cybersecurity Literature and Practice

Taking cognizance of the lack of cybersecurity-specific role's focus in studies, cybersecurity scholars have contributed to literature from multiple perspectives. First, studies have articulated the substantial, yet, latent rising costs to businesses in responding to cyber threats by only focusing on technology-centric solutions [4, 17]. Second, championing the importance of incorporating human factors in the cybersecurity practice, numerous studies have advocated for leveraging cognitive scientists and human factor experts to conduct assessments on human performance on aspects such as cognitive overload and technostress, in a technology-driven work environment [18, 19]. Third, and arguably, most critical to this endeavor is the concerted effort to define and establish a human-centered cybersecurity (HCC) framework. The HCC framework places humans at the center of cybersecurity and information security practices, design aspects, and technology integration to reduce behavioral-centric risks by accounting for psychological efforts and impacts [20, 21].

Despite this multi-pronged approach, there exist several key gaps in the current state of affairs. The lack of academia-industry collaboration has continued to inhibit just-in-time and symbiotic exchange of insights to holistically inform the HCC domain [22]. This has resulted in research output that has low likelihood of practitioner uptake, and well-intentioned practitioners lacking insights on how to implement HCC principles in an organizational setting. Second, and a more profound omission has been the failure to borrow valuable lessons-learned and time-tested practices from other domains, which have demonstrated relative maturation. Specifically, the Healthcare profession has adopted a socio-technical systems perspective in dealing with human aspects. Socio-technical systems are broadly defined as *involving the interaction of computing systems and human beings, in ways that* either cannot be separated or are thought to be inappropriate to separate [23]. The underlying premise of socio-technical thinking is that systems design should be a process that considers both social *and* technical factors that influence the functionality and usage of computing systems [24].

Academic research continues to evolve from viewing cybersecurity as purely a technical problem to one in which a broader socio-technical lens is adopted to investigate cybersecurity behaviors, risks, and culture [22, 25]. However, more needs to be done. A holistic solution to address human aspects requires a that the science and technology deployed to protect and defend our information and critical infrastructure consider human, social, organizational, economic, and technical factors, as well as the complex interaction among them, in the creation, maintenance, and operation of our systems and infrastructure [26]. These apertures in the current-state collectively served as the driving motivation for convening a multidisciplinary workshop at the 26th International Conference on Human-Computer Interaction.

### 1.3    Workshop Objectives

Experts across academia and industry represented by academic researchers, cybersecurity leadership, behavioral psychologists, and human factors engineers discussed challenges surrounding the topic of burnout in the cybersecurity profession, with the broader goal of laying a foundational roadmap for joint academy-industry collaborative research. Table 2 outlines the primary workshop themes.

**Table 2.** Workshop Objectives by Theme

---

Theme & Discussion Questions

---

### 1. State-of-the-art academic literature on burnout in the cybersecurity profession

- What definitions of burnout have been used in cybersecurity literature?
- What interventions and organizational policies have been discussed or implemented to mitigate burnout among cybersecurity workers, and with what success?
- What gaps exist in the current literature regarding burnout in cybersecurity, and what areas require further empirical research?

### 2. Antecedents of burnout

- What are the causes of burnout in cybersecurity?
- How does organizational culture within the cybersecurity industry influence the development and management of burnout?
- How do the stressors unique to cybersecurity, such as the persistent threat environment and the rapid pace of technological change, exacerbate the risk of burnout?

### 3. Manifestation of burnout

- How do cybersecurity professionals recognize or identify the onset of burnout?
- How do signs of burnout manifest at the individual contributor, mid-level manager, and senior leader levels?

### 4. Impact and Prevention of burnout

- How do cybersecurity professionals deal with burnout today?
- What are the impacts of burnout at an individual level?
- What are the impacts of burnout at an organizational level?
- What preventative strategies effectively reduce the risk or impact of burnout among cybersecurity professionals?
- How have other socio-technical professions approached burnout?

---

The rest of the paper is organized as follows. We first present the state-of-the-art literature on burnout, followed by a deeper discussion on the antecedents, manifestation and impact of burnout amongst cybersecurity professionals. Central to this examination is an empirical study conducted with former tactical cyber operators from the National Security Agency (NSA) in the United States. Next, we propose a roadmap comprising of a comprehensive set of strategies for mitigating burnout among cybersecurity professionals, inclusive of good practices and lessons-learned from the Healthcare domain. We conclude with a collective call-to-action for academia and industry practitioners to advance this time-sensitive and critical endeavor.

## 2      What is Burnout

### 2.1      Burnout Definition

The World Health Organization characterizes burnout as an occupational phenomenon, highlighting its increasing prevalence within cybersecurity [6]. Literature offers varying yet closely correlated definitions of burnout. Malasch et al. [44] defined burnout as chronic response to job stressors characterized by emotional exhaustion, cynicism, and reduced self-efficacy, while other scholars have characterized burnout as emotional weariness and exhaustion, depersonalization, lacking competence, and a detached attitude about your work [45], noting that burnout is not an individual-based issue but is caused by the workplace climate and environment [46].

## 3      Antecedents of Burnout

In cybersecurity, human performance challenges (stress, fatigue, distractions, and burnout) are not just prevalent, but they also pose a significant threat. Numerous factors further intensify these challenges, including remote working conditions, disruptions, significant life events, and a persistent threat landscape [4][47]. A study by Cunningham [48] highlighted that chronic stress, anxiety, frequent interruptions, and burnout are critical human performance issues that demand immediate and proactive leadership engagement. The compounding factors, such as remote work, the dynamic nature of cybersecurity threats, constant change, life stressors, and daily routines, severely hamper human performance. The overreliance on a technology-led cycle only serves to exacerbate these issues.

The relentless integration of advanced technologies has increased human and processual components, resulting in heightened operational complexity [49]. This complexity often diminishes visibility and control over processes, undermining cybersecurity measures' efficacy [49]. As a result, this erosion of process oversight leads to a decline in human performance and amplifies the likelihood of cybersecurity vulnerabilities [49]. The intricate interplay between escalating technological integration and human factors necessitates not just a different approach but a more nuanced one. This approach should not just balance technological advancements with robust human-centered strategies but also delve into the depths of these issues and address them comprehensively. Most organizations are not implementing human-centered strategies, significantly contributing to human performance issues.

It is widely recognized that human vulnerabilities represent the most substantial risk to cybersecurity programs [50]; however, many business leaders fail to address these human performance factors adequately. The continued prevalence of these issues stems from a lack of comprehensive education on human factors within cybersecurity. The glaring issue within the cybersecurity sector continues to be the absence of effective preventive and mitigating strategies to tackle the pervasive problems of fatigue, stress, and burnout. Organizations are currently inadequate in their proactive engagement to circumvent the genesis of burnout, failing to address the underlying causes before they manifest. Effective prevention necessitates addressing the sustained exposure to chronic stress and fatigue.

Compounding this problem is a significant dearth of research focused on strategies to forestall burnout, specifically in the cybersecurity domain. Until concerted efforts are made to mitigate early indicators such as stress and fatigue, burnout is poised to continue its detrimental impact, further exacerbated by the relentless pace and demands of technology-driven cycles. In summary, the core of the issue extends beyond mere symptoms of occupational stress; it lies in the systemic failure to preemptively manage and mitigate significant exposure to these stressors, thereby averting burnout and cognitive decline.

## 4        Manifestation and Impacts of Burnout – A United States National Security Agency Case Study

In this section, we present an empirical study conducted at the United States National Security Agency (NSA) on the manifestation of burnout and its consequences on professionals in the tactical cyber operator role. The study's findings serve as a vital reminder for both the public and private sector on the need for emphasizing the Human aspects of cybersecurity.

The NSA is a United States government agency that collects and analyzes foreign intelligence to protect national security. NSA tactical cyber operators are a specialized workforce that are highly trained, highly skilled, and highly valued [42]. They perform extraordinary tasks under stressful conditions in an environment that is complex, elevated risk, and has limitations on compensation. Such work and conditions have an impact on the workforce. These concerns are echoed in the commercial sector related to stress and burnout. Burnout has led to high turnover, both inside the government and in the commercial sector. This turnover is expensive for organizations and puts critical cyber missions at risk from reduced expertise and readiness. A belief among some organizational leaders is that cyber operators leave their jobs for reasons related to compensation; however, anecdotal evidence led the NSA to conduct research to uncover true reasons why people left their job.

The NSA conducted a study that investigated the reasons why people leave work in cyber operations. The research team included a clinical psychologist, a former operator, and a human-factors researcher. Semi-structured interviews with 10 former tactical cyber operators who left government service revealed themes and perspectives about their prior work. Between June and August of 2023, the team interviewed 10 former NSA operators to understand their reasons for leaving government service. These operators had an average of 8.4 years of active status between 2006 and 2023. Two operators were certified as Apprentices, five were Journeymen, and three were Master operators. All but one of these operators served during the height of the counter-terrorism mission. Interviews focused on the experience, pros and cons of operations, factors that led to them leaving, and general regrets. Interviews were conducted in a way that was cognizant of triggers that would lead to emotional distress. None of the participants held an active security clearance or worked at NSA at the time of the interviews.

The empirical findings reveal intense passion for the mission while also highlighting high levels of stress that negatively impacted physical and emotional health. With the distance of time, participants understood the toll of their work. This study contributes to the body of research on cybersecurity information workers by providing in-depth insights into the experiences of cyber operators, offering suggestions for the long-term health and safety of the workforce, and highlighting the need for revised approaches to hiring and retention [43].

### 4.1    Research findings

Most operators left because of natural career progression. The desire for benefits that would support a better work-life balance (normal schedule, working from home) were the second most common reasons for leaving. None of the operators cited pay as a reason for leaving.

The desire for better work-life balance came from years of job-related stress that led to burnout. Work-related trauma, a stressful work environment, and long hours were contributing factors to the mental distress and strained relationships that led to burnout.

Some of the operators may have extended their careers at NSA if they had access to career mentoring, soft benefits to help work-life balance, or access to an embedded psychologist.

### 4.2    Compensation and Job Satisfaction as Indicators for Attrition

**Pros and cons of working at NSA**. The most common pros of working at the NSA were the challenging work and the mission. The most common cons were the burn out associated with the operations environment.

All but one of the operators served during the height of the counter-terrorism mission. While that mission is over, the United States is in the midst of supporting a kinetic war in Russia and preparing for conflict with China. The source of trauma from violent material and resulting emotional distress is similar.

**Love for the mission**. Eight of 10 participants spoke enthusiastically about having meaningful work with a sense of purpose and satisfaction. They also said that they "missed" the uniquely-government mission after they left. Upon reflection, former operators appreciated the privilege of the work they did and the sensitive, secret, and exciting nature of the work. Seven operators described their love of puzzles and technical challenges. Three operators described the attraction to the mission as "addicting" or "riding the dragon." This love for the challenging mission is probably a significant reason that none of the operators had negative feelings about their time as an operator, even if they experienced serious and lingering mental health effects, such as post-traumatic stress disorder (PTSD) and poor sleep.

Love for the unique challenges of the mission influenced operators' desires to return to the NSA. Additionally, despite their burnout and leaving the intelligence community completely, the lingering positive feelings of their NSA experiences reduces the likelihood that these operators pose a counterintelligence risk.

**Burnout and mental health**. The overwhelming love of the mission may have masked negative environmental factors and personal impacts, or had extended operators' time beyond when they should have left without negative impacts to themselves, the mission, and the offices they were assigned. Prior research in stress and cyber operations found that performance is an utmost priority for operators despite impacts to mental health. As a result, they could both love their jobs while pushing themselves to burnout. Special pay scales could exacerbate this danger if an operator feels unable to escape "golden handcuffs" and explore other career paths at NSA. Interestingly, several operators initially stated they would come back to the NSA in a heartbeat, but then quickly corrected themselves and cited that their spouses would never allow it. The most common negative psychological symptoms were sleep disturbance and deprivation. Six participants mentioned impacts on sleep. As one operator put it, their "work-life balance was out of whack." Several operators noted that sleep improved significantly after they left. However, other operators continued to suffer from sleep issues even years after leaving NSA.

Burnout is dangerous because of its risks to mission and operators. Tired, stressed operators will not make optimal decisions during high-risk ops. Tired, stressed operators are at risk of car accidents, sleep deprivation, and relationship issues. Finding ways to rebalance the work-life relationship would reduce risk to mission while extending the shelf life of an operator.

**Life outside of the NSA**. The most noted soft benefits to working outside of the NSA was the flexibility to move and work from home to support better work-life balance. The most common complaint about leaving the NSA was that the work is not as interesting. Interestingly, no one cited Pay as the primary benefit for staying, although a few acknowledged that the extra money from incentives was nice.

Soft benefits will likely have a bigger impact on retention than pay, but they are also the most difficult to manage equitably. Incentive pay should not be taken away, but it is important to acknowledge its limited impact on attrition.

### 4.3 Recommendations

Based on the research findings, the research team proposed four recommendations.

**Mandatory rest periods**. Fatigue was a major influence on stress and burnout. "Crew rest" would help ensure that operators remain safe and competent during operations. This could be implemented as a minimum number of off-duty non-operational hours before the next operation. Additionally, mandatory breaks during an operation could help manage fatigue. Team-based ops tradecraft could safely extend the time of an operation and allow operators to take intermediary breaks. This would not replace between-op crew rest but may reduce the minimum number of hours between operations.

**Retiring from operations**. Organizations should consider establishing a maximum tenure for which an operator may conduct active cyber ops. However, this could include an off-ramp where experienced operators continue to be on stand-by for a certain period after rotating out of active ops. This approach keeps experienced operators involved though knowledge transfer by reassigning them as trainers, mentors, and certifiers.

**Career progression path**. Expanding incentive pay to approved diversity tours may help operators find additional paths for their careers. Operators can perhaps retain their incentive pay in relevant but non-incentive pay positions for up to 3 years before returning to an incentive position. Diversity tours help spread knowledge while providing learning experiences and breaks from the operational environment. However, more mentoring on what a post-operational career path looks like would be needed.

**Lifetime helpline**. Establish resources that are easily accessible for people who leave government service, such as a phone number or email address for questions about classification. This could have considerable impacts on managing risk and possible re-recruitment of former operators.

These empirical findings reveal intense passion for the mission while also highlighting prominent levels of stress that negatively impacted physical and emotional health. With the distance of time, participants understood the toll of their work. This study contributes to the body of research on security information workers by providing in-depth insights into the experiences of cyber operators, offering suggestions for the long-term health and safety of the workforce, and highlighting the need for revised approaches to hiring and retention [43].

## 5      A Comparative Review of Burnout in the Healthcare Profession

Valuable lessons learned from the NSA case study underscore the need for a comprehensive approach to address burnout in the cybersecurity profession. Such an endeavor can benefit from borrowing good practices in allied socio-technical domains such as healthcare, where the well-being of healthcare providers and the influence of their mental health and wellness on patient safety have long been studied [27]. The discussion presented next highlights the many parallels between healthcare and cybersecurity professions which require due examination in the context of burnout.

Healthcare is a complex and dynamic socio-technical system in which groups of people cooperate for patient care and are faced with numerous contingencies that cannot be fully anticipated [28]. In addition, healthcare organizations may have conflicting values and objectives (e.g., financial objectives and humanitarian objectives) and are increasingly diverse and subject to considerable pressures and changes [29]. Similarly, the cybersecurity professional is constantly tasked with balancing the often conflicting needs of risk vs. safety, short-term vs. long-term focus, operational efficiency vs. security measures, ethical considerations vs. practical constraints and, most importantly, navigating the fine line between the human factor and systemic challenges. These similarities warrant further investigative questions:

- What are the innate differences on how occupational burnout has been approached in healthcare vs. cybersecurity professions?
- Are there valuable lessons learned for the cybersecurity profession from the breadth and depth of healthcare academic literature?
- What is a forward-looking interdisciplinary agenda that can be developed to further research from a sociotechnical perspective?

## 5.1    Healthcare Burnout Studies by Specialization

Healthcare literature dating back several decades bears evidence of the extant exploration of burnout among physicians, nurses and healthcare staff. A 1975 study by Freudenberg explored the burnout syndrome in a clinical setting and not only forwarded a definition of burnout but also discussed its symptoms and preventive measures by healthcare professional role.  In addition, studies have analyzed burnout by healthcare specialty. For example, *Eelen et al.* [30] have analyzed the prevalence of burnout among oncology professionals. *Kerlin et al.* [31]. Reviewed the epidemiology of burnout syndrome in the intensive care unit (ICU) and the impact it can have on clinicians, patients, and the health service.

## 5.2    Healthcare Burnout Studies by Geography

Healthcare studies have also focused the burnout phenomenon by continents and countries. *Bria et al.* [32] conducted a systematic review of burnout risk factors among European healthcare professionals. *Dubale, et al.* [33] studied burnout among healthcare providers in sub-Saharan Africa. *Mercado et al.* [34] conducted a cross-sectional analysis of factors associated with stress, burnout and turnover intention among healthcare workers during the COVID-19 pandemic in the United States. *Tan et al.* [35] studied the prevalence of burnout among various groups of healthcare professionals in Singapore through a quantitative survey using the Maslach Burnout Inventory to measure three categories of burnout: emotional exhaustion, de-personalization and personal accomplishment. Studies have also compared if burnout differences across countries were statistically significant. For instance, one study compared burnout levels between Italian and Dutch healthcare professionals and concluded that Italian healthcare professionals experienced higher burnout scores [36]. The authors explained differences as a consequence of unfavorable job characteristics, like high work and time pressure or high physical demands.

### 5.3     Technology as an Antecedent to Healthcare Professional's Burnout

Of note, is the focus on technology's impact on healthcare professional's well-being. This is demonstrated in the study by *Ehrenfeld & Wanderer* [37], where they investigated the impact of electronic health records (EHR) on physician burnout. Their study found numerous factors including poor usability, incongruent workflows, and the addition of clerical tasks to physician documentation requirements as ongoing concerns with EHR adoption. Their study highlighted that continued use of EHRs decreased professional satisfaction, increased burnout, and ultimately, the physician's likelihood to reduce or leave clinical practice. *Privitera et al.* [38] surveyed 1,048 physicians in New York to explore how time spent on EHRs at home affected physician's job satisfaction, job stress and burnout. Conclusions from their study identified physicians' moderately high to excessive time spent on EHRs at home, did not significantly affect job satisfaction. In addition, EHR use at home significantly increased their odds of experiencing job stress by 50% and burnout by 46%. Moreover, physician's length and degree of documentation requirements and extension of work life into home by means of e-mail, completion of records and phone calls significantly correlated to decreased job satisfaction and increased job stress and likelihood of burnout.

By contrast, the study of human factors in cybersecurity is an emerging phenomenon. Far fewer studies have explored the burnout syndrome by cybersecurity role (for example, Ethical Hacker and Forensic Analyst), compared to healthcare roles. More recently, researchers have begun to explore burnout among cybersecurity incident responders [39] and Chief Information Security Officers [1]. Secondly, there is no evidence of empirical studies comparing burnout among cybersecurity professionals across countries and continents. Lastly, and perhaps most critically, there is a dearth of empirical studies investigating the impact of technology on burnout among cybersecurity professionals.

## 6     Roadmap For Future Research to Address Burnout

Tackling burnout is crucial for maintaining cybersecurity expertise and strengthening the resilience of digital defenses. However, creating a comprehensive program to address burnout will require collaboration across industry, academia, and both public and private sectors to drive lasting change. We highlight several key areas to guide further development and research in this field.

### 6.1     Fatigue Management Programs

A fundamental element of any effort to mitigate fatigue is the implementation of a Fatigue Risk Management System (FRMS) [51]. The Federal Aviation Administration of the United States defines an FRMS as a "scientifically based, data-driven process and systematic method used to monitor and manage fatigue risks associated with fatigue-related errors continuously" [52]. While an FRMS can be an integral component of an organization's Safety Management System (SMS), it is not a mandatory inclusion. This system emphasizes proactively identifying and mitigating fatigue risks to enhance overall safety, with a strong focus on reducing organizational risk [51].

The cybersecurity community should work to implement Cybersecurity Fatigue Management Programs to identify, prevent, reduce, and mitigate factors that increase fatigue. One potential goal of a Cybersecurity Fatigue Management Program is to prevent cybersecurity professionals from reaching burnout. Another goal is to foster discourse on strategies to avoid burnout through mitigating actions executed via a Cybersecurity Fatigue Management Program. Examining and replicating the anti-fatigue solutions leveraged by other sociotechnical domains is essential.

Fatigue management in cybersecurity is paramount due to the critical role of vigilance and sustained cognitive function in maintaining secure systems. Cybersecurity professionals are often required to monitor networks, analyze threats, and respond to incidents around the clock, leading to prolonged periods of intense concentration and stress. Without proper fatigue management, these individuals are at a heightened risk of errors, slower reaction times, and diminished decision-making abilities, all of which can compromise an organization's security posture. Effective fatigue management practices, such as structured work-rest schedules, regular breaks, and promoting a healthy work-life balance, ensure that cybersecurity professionals remain alert, focused, and capable of effectively mitigating potential threats. By prioritizing the well-being of these key personnel, organizations can enhance their overall cybersecurity resilience and reduce the likelihood of successful cyberattacks.

## 6.2     Longitudinal Studies that Inform Intervention Strategies

The NSA case study highlights the need for more in-depth longitudinal inquiries that track the cybersecurity professional's career trajectory and burnout patterns over time. Such analyses can both provide insights on the effectiveness of introducing burnout prevention and mitigation strategies, as well as help build a comparative profile for interventions specific to cybersecurity roles. In addition, such studies can help inform the development of tailored assessment scales to self-evaluate burnout by role.

## 6.3     Categorization of Stressors

As we proceed to develop socio-technical solutions, we can borrow from other disciplines and break the large and complex problems into smaller addressable segments. With years of experience in delivering software projects across many industries, addressing complex problems by breaking the work into specific work packages, developing user stories, or chunking the work, it is reasonable to offer a similar approach to the cybersecurity burnout problem. With respect to the thematic burnout topics discussed at the workshop, one such viable option is categorizing stressors into job content stressors and job context stressors.

This categorization can offer scope parameters and language for practitioners and researchers [40]. Borrowing from literature studying police officers, job content refers to stressors related to those specifically related to role tasks, while job context refers to stressors related to the organization; operational and organizational, respectively [41]. When applied to cybersecurity, job content means stressors that arise from using cybersecurity tools while responding to an incident, and job context means stressors that arise from poor leadership [40]. Creating frameworks that leverage the dual job content-job context taxonomy can serve as beneficial tools for organizations to navigate the complex cybersecurity burnout landscape.

### 6.4    Incorporating Lessons-Learned from the Healthcare Profession

By learning from the healthcare profession's multidimensional approach to address burnout, the cybersecurity profession can take proactive steps to improve the well-being of its professionals.

- **Acknowledge and Elevate the Discourse on Burnout.** The healthcare profession has long recognized burnout as a critical issue within its ecosystem. For cybersecurity, it is essential to similarly acknowledge burnout as a pervasive challenge affecting professionals across the industry. Beyond traditional academic publications, cybersecurity academics and practitioners should explore avenues like industry forums, conferences, webinars, and online platforms to share experiences, raise awareness, and engage in meaningful conversations about burnout
- **Recognize Critical Points of Impact, Like the Pandemic.** In healthcare, major events like the COVID-19 pandemic have led to heightened awareness of burnout among healthcare professionals. Cybersecurity practitioners and academics alike can learn from this by considering how external crises, such as global cyberattacks or the shift to remote work, amplify burnout in the industry. Such occasions can serve as opportunity to conduct post-event studies or surveys to understand the stressors faced by professionals and the unique challenges they endure during crises
- **Treat Technology as a Double-Edged Sword.** Technology is often seen as a "villain" contributing to the healthcare professional's burnout due to overreliance on digital systems and lack of integration. Similarly, in cybersecurity, technology can exacerbate burnout by creating a constant "always-on" environment, with professionals being overwhelmed by too many tools, alerts, and data to manage. It is imperative to take this into consideration in the design and development of cybersecurity tools and technologies. Cybersecurity professionals should also consider how technology can contribute to burnout by leading to information overload, alert fatigue, or tool inefficiencies. Improving the usability of security tools, enhancing automation to reduce manual workloads, and creating better integrations can help reduce stress levels.
- **Holistic Approach to Well-Being.** Healthcare has recognized the importance of an integrated approach to well-being, addressing mental, emotional, and physical health to mitigate burnout. Similarly, cybersecurity organizations should adopt a compre-hensive approach to professional well-being, including mental health support, flexible work environments, and adequate time for rest. Healthcare workflow practices such as formal shift turnovers, and role rotations can provide opportunities for the cybersecurity profession in adopting similar welfare practices.

- **Global Collaboration to Tackle Burnout.** As evidenced by literature studies, tackling burnout in healthcare has become a global priority. Global collaborations can similarly elevate the focus on cybersecurity burnout, with professionals and researchers sharing best practices, research findings, and interventions across geographic boundaries. International conferences, joint research projects, and online platforms dedicated to cybersecurity well-being can create a global effort to address burnout. Collaboration can also help standardize burnout prevention programs and tools across organizations worldwide.
- **Continuous Research and Evaluation of Burnout.** The healthcare sector consistently evaluates burnout through ongoing research and surveys to understand its impact and develop effective interventions. The cybersecurity profession should similarly invest in periodic studies and surveys to track the prevalence of burnout, identify its root causes, and assess the effectiveness of implemented solutions. Establishing regular burnout assessments within cybersecurity teams can help evaluate mental health trends, stress levels, and risk factors. Moreover, this can present collaboration avenues with academic institutions to conduct longitudinal studies and track progress in mitigating burnout over time.

## 7    Conclusion

In conclusion, addressing burnout in cybersecurity is not merely an issue of employee well-being, but a critical factor in maintaining robust defenses. Cybersecurity professionals and organizations globally are at risk. The time-sensitive and action-oriented outcomes discussed in the workshop are no longer optional, but imperative. By synthesizing insights from diverse sources and proposing concrete mitigation strategies, we catalyze a shift in workforce sustainability. As the threat landscape evolves, ensuring long-term resilience of human cybersecurity resources is paramount. Our work is thus a collective call-to-action for researchers, organizations, and policymakers alike to safeguard the very defenders of our digital systems.

# References

1. Reeves, A., Pattinson, M. and Butavicius, M., 2023. Is Your CISO Burnt Out Yet? Examining Demographic Differences in Workplace Burnout Amongst Cyber Security Professionals. In: *International Symposium on Human Aspects of Information Security and Assurance*. Cham, Switzerland: Springer Nature, pp. 225–236.

2. Devo, 2024. The Modern CISO: An Essential Guide for CISO Success. Available at: https://www.devo.com/resources/ebook/the-modern-ciso-an-essential-guide-for-ciso-success/ [Accessed 21 September 2024].

3. Budge, J., Roberts, J., Shey, H. and Levine, D., 2023. We need to talk more about burnout in Cybersecurity. *Forrester*, 14 February. Available at: https://www.forrester.com/blogs/we-need-to-talk-more-about-burnout-in-cybersecurity/.

4. Nobles, C., 2022. Stress, burnout, and security fatigue in cybersecurity: A human factors problem. *HOLISTICA – Journal of Business and Public Administration*, 13(1), pp. 49–72. Available at: https://doi.org/10.2478/hjbpa-2022-0003.

5. Platsis, G., 2019. The Human Factor: Cyber Security's Greatest Challenge. In: *Cyber Law, Privacy, and Security: Concepts, Methodologies, Tools, and Applications*. IGI Global, pp. 1–19.

6. World Health Organization, 2019. QD85 Burnout. In: *International Statistical Classification of Diseases and Related Health Problems* (11th ed.). Available at: https://icd.who.int/browse11/l-m/en#/http://id.who.int/icd/entity/129180281.

7. Gupta, V., Rangarajan, A. and Nobles, C., 2024. Burnout in the Cybersecurity Profession: A Scoping Review. *MWAIS 2024 Proceedings*, 20. Available at: https://aisel.aisnet.org/mwais2024/20.

8. King, R.C. and Sethi, V., 1997. The moderating effect of organizational commitment on burnout in information systems professionals. *European Journal of Information Systems*, 6(2), pp. 86–96.

9. Moore, J.E., 2000. One road to turnover: An examination of work exhaustion in technology professionals. *MIS Quarterly*, pp. 141–168.

10. Pawlowski, S., Kaganer, E. and Cater III, J., 2004. Mapping perceptions of burnout in the information technology profession: A study using social representations theory. *ICIS 2004 Proceedings*, p. 73.

11. Salanova, M., Peiró, J.M. and Schaufeli, W.B., 2002. Self-efficacy specificity and burnout among information technology workers: An extension of the job demand-control model. *European Journal of Work and Organizational Psychology*, 11(1), pp. 1–25.

12. Sethi, V., Barrier, T. and King, R.C., 1999. An examination of the correlates of burnout in information systems professionals. *Information Resources Management Journal (IRMJ)*, 12(3), pp. 5–13.

13. Shih, S.P., Jiang, J.J., Klein, G. and Wang, E., 2013. Job burnout of the information technology worker: Work exhaustion, depersonalization, and personal accomplishment. *Information & Management*, 50(7), pp. 582–589.

14. Brooks, N.G., Hardgrave, B.C., O'Leary-Kelly, A.M., McKinney, V. and Wilson, D.D., 2015. Identifying with the information technology profession: Implications for turnover of IT professionals. *ACM SIGMIS Database: The Database for Advances in Information Systems*, 46(1), pp. 8–23.

15. Rutner, P., Riemenschneider, C., O'Leary-Kelly, A. and Hardgrave, B., 2011. Work exhaustion in information technology professionals: The impact of emotion labor. *ACM SIGMIS Database: The DATABASE for Advances in Information Systems*, 42(1), pp. 102–120.

16. Kim, S., 2005. Factors affecting state government information technology employee turnover intentions. *The American Review of Public Administration*, 35(2), pp. 137–156.
17. Neely, L., 2017. Threat landscape survey: Users on the front line. *SANS Institute*. Available at: https://www.sans.org/reading-room/whitepapers/threats/2017-threat-landscape-survey-users-frontline-37910.
18. National Security Agency, 2015. Science of Security (SoS) Initiative Annual Report 2015. Available at: http://cps-vo.org/sos/annualreport2015.
19. Nobles, C., 2015. Exploring pilots' experiences of integrating technologically advanced aircraft within general aviation: A case study. *Northcentral University*.
20. Bureau, S., 2018. Human-centered cybersecurity: A new approach to securing networks. *Rochester Institute of Technology Research Report*, Fall/Winter 2018.
21. ForcePoint Security Labs, 2018. 2018 Security Predictions. Available at: https://www.forcepoint.com/sites/default/files/resources/files/report_2018_security_predictions_en.pdf.
22. Haney, J., Cunningham, C. and Furman, S.M., 2024. Towards Integrating Human-Centered Cybersecurity Research Into Practice: A Practitioner Survey. In: *Symposium on Usable Security and Privacy (USEC)*, 2024.
23. Pallister, J. and Law, J., 2006. *A Dictionary of Business and Management*.
24. Baxter, G. and Sommerville, I., 2011. Socio-technical systems: From design methods to systems engineering. *Interacting with Computers*, 23(1), pp. 4–17.
25. McEvoy, T.R. and Kowalski, S.J., 2019. Deriving cyber security risks from human and organizational factors – A socio-technical approach. *Complex Systems Informatics and Modeling Quarterly*, no. 18, pp. 47–64.
26. Sociotechnical Cybersecurity, Computing Community Consortium, n.d. Available at: https://cra.org/ccc/visioning/visioning-activities/2016-activities/sociotechnical-cybersecurity/.
27. Bridgeman, P.J., Bridgeman, M.B. and Barone, J., 2018. Burnout syndrome among healthcare professionals. *The Bulletin of the American Society of Hospital Pharmacists*, 75(3), pp. 147–152.
28. Carayon, P., Bass, E.J., Bellandi, T., Gurses, A.P., Hallbeck, M.S. and Mollo, V., 2011. Sociotechnical systems analysis in health care: A research agenda. *IIE Transactions on Healthcare Systems Engineering*, 1(3), pp. 145–160.
29. Effken, J.A., 2002. Different lenses, improved outcomes: A new approach to the analysis and design of healthcare information systems. *International Journal of Medical Informatics*, 65(1), pp. 59–74.
30. Eelen, S., Bauwens, S., Baillon, C., Distelmans, W., Jacobs, E. and Verzelen, A., 2014. The prevalence of burnout among oncology professionals: Oncologists are at risk of developing burnout. *Psycho-Oncology*, 23(12), pp. 1415–1422.
31. Kerlin, M.P., McPeake, J. and Mikkelsen, M.E., 2020. Burnout and joy in the profession of critical care medicine. In: *Annual Update in Intensive Care and Emergency Medicine 2020*, pp. 633–642.
32. Bria, M., Baban, A. and Dumitrascu, D.L., 2012. Systematic review of burnout risk factors among European healthcare professionals. *Cognition, Brain, Behavior: An Interdisciplinary Journal*, 16(3), pp. 423–452.
33. Dubale, B.W., Friedman, L.E., Chemali, Z., Denninger, J.W., Mehta, D.H., Alem, A., Gelaye, B., 2019. Systematic review of burnout among healthcare providers in sub-Saharan Africa. *BMC Public Health*, 19, pp. 1–20.

34.  Mercado, M., Wachter, K., Schuster, R.C., Mathis, C.M., Johnson, E., Davis, O.I. and Johnson-Agbakwu, C.E., 2022. A cross-sectional analysis of factors associated with stress, burnout, and turnover intention among healthcare workers during the COVID-19 pandemic in the United States. *Health & Social Care in the Community*, 30(5), pp. e2690–e2701.

35.  Tan, K.H., Lim, B.L., Foo, Z., Tang, J.Y., Sim, M., Lee, P.T. & Fong, K.Y., 2022. Prevalence of burnout among healthcare professionals in Singapore. *Annals of the Academy of Medicine Singapore*, 51(7), pp.409–416.

36.  Pisanti, R., van der Doef, M., Maes, S., Lazzari, D. & Bertini, M., 2011. Job characteristics, organizational conditions, and distress/well-being among Italian and Dutch nurses: A cross-sectional comparison. *International Journal of Nursing Studies*, 48, pp.829–837. https://doi.org/10.1016/j.ijnurstu.2010.12.006.

37.  Ehrenfeld, J.M. and Wanderer, J.P., 2018. Technology as friend or foe? Do electronic health records increase burnout? *Current Opinion in Anesthesiology*, 31(3), pp. 357–360.

38.  Privitera, M.R., Atallah, F., Dowling, F., Gomez-DiCesare, C., Hengerer, A., Arnhart, K. & Staz, M., 2018. Physicians' electronic health records use at home, job satisfaction, job stress and burnout. *Journal of Hospital Administration*, 7(4), pp.52–59.

39.  Nepal, S., Hernandez, J., Lewis, R., Chaudhry, A., Houck, B., Knudsen, E., Czerwinski, M., 2024. Burnout in cybersecurity incident responders: Exploring the factors that light the fire. *Proceedings of the ACM on Human-Computer Interaction*, 8(CSCW1), pp. 1–35.

40.  Hollis, T.R., 2023. All Quiet on The Digital Front: The Unseen Psychological Impacts on Cybersecurity First Responders. University of South Florida.

41.  Shane, J.M., 2010. Organizational stressors and police performance. Journal of criminal justice, 38(4), pp.807-818.

42.  Dykstra, J. and Paul, C.L. "Cyber Operations Stress Survey (COSS): Studying fatigue, frustration, and cognitive workload in cybersecurity operations." 11th USENIX Workshop on Cyber Security Experimentation and Test (CSET 18). 2018.

43.  Paul, C.L. and Dykstra, J., 2017. Understanding operator fatigue, frustration, and cognitive workload in tactical cybersecurity operations. *Journal of Information Warfare*, 16(2), pp.1-11.

44.  Maslach, C., Schaufeli, W.B. and Leiter, M.P., 2001. Job burnout. *Annual review of psychology*, 52(1), pp.397-422.

45.  Pham, H.C., Brennan, L. and Furnell, S., 2019. Information security burnout: Identification of sources and mitigating factors from security demands and resources. *Journal of Information Security and Applications*, 46, pp.96-107.

46.  Maslach, C. and Leiter, M.P., 2005. Reversing burnout. *Standford Social Innovation Review*, 34(4), pp.43-49.

47.  Robinson, N. (2023). Human factors security engineering: The future of cybersecurity teams. *EDPACS*, 67(5), 1-17.

48.  Cunningham, M. (2021). *"Tiny crimes" – How minor mistakes when remote working could lead to significant cybersecurity breaches (Part 1).* [online] Forcepoint.com. Available at: https://www.forcepoint.com/blog/x-labs/minor-mistakes-major-breaches-pt-1

49.  Stallbaum, S.A.W., Dean Hamilton, and Scott (2020). *The Unaddressed Gap in Cybersecurity: Human Performance.* [online] MIT Sloan Management Review. Available at: https://sloanreview.mit.edu/article/the-unaddressed-gap-in-cybersecurity-human-performance/.

50.  Moustafa, A. A., Bello, A., & Maurushat, A. (2021). The role of user behaviour in improving cyber security management. *Frontiers in Psychology*, *12*, 561011.

51.  Bendak, S. and Rashid, H.S., 2020. Fatigue in aviation: A systematic review of the literature. *International Journal of Industrial Ergonomics*, *76*, p.102928.

52. Faa.gov. (2010). *AC 120-100 - Basics of Aviation Fatigue.* [online] Available at: https://www.faa.gov/regulations_policies/advisory_circulars/index.cfm/go/document.information/documentID/244560.